\documentclass[10pt]{elsart}
\usepackage{amssymb}
\usepackage{amsmath}
\usepackage{graphicx}
\journal{Physics Letters A}
\begin{document}
\begin{frontmatter}

\title{A class of ($\ell$-dependent) potentials with the same number of ($\ell$-wave) bound states}

\author{Fabian Brau\thanksref{fnrs}}
\thanks[fnrs]{FNRS Postdoctoral Researcher}
\address{Service de Physique G\'en\'erale et de Physique des Particules El\'ementaires, Groupe de Physique Nucl\'eaire Th\'eorique, Universit\'e de Mons-Hainaut, Mons, Belgique}
\ead{fabian.brau@umh.ac.be}

\author{Francesco Calogero}
\address{Dipartimento di Fisica, Universit\`a di Roma ``La Sapienza" and Istituto Nazionale di Fisica Nucleare, Sezione di Roma, Rome, Italy}
\ead{francesco.calogero@roma1.infn.it}

\date{\today}

\begin{abstract}
We introduce and investigate the class of central potentials
\begin{displaymath}
V_{\text{CIC}}(g^{2},\mu^{2},\ell,R;r)=-\frac{g^{2}}{R^{2}}\,
\left(\frac{r}{R}\right)^{4\ell}\,\left\{ \left[ 1+\left(\frac{1}{2\ell+1}\right)
\left(\frac{r}{R}\right)^{2\ell+1}\right]^{2}-1+\mu^{2}\right\}^{-2},
\end{displaymath}
which possess, in the context of nonrelativistic quantum mechanics, a number
of $\ell$-wave bound states given by the ($\ell$-independent !) formula 
\begin{displaymath}
N_{\ell}^{\text{(CIC)}}\left(g^{2},\mu ^{2}\right) =\left\{\left\{ 
\frac{1}{\pi}\sqrt{g^{2}+\mu^{2}-1}\,\left(\sqrt{\mu^{2}-1}\right)^{-1}
\arctan\left(\sqrt{\mu^{2}-1}\right) \right\} \right\}.
\end{displaymath}
Here $g$ and $\mu$ are two arbitrary real parameters, $\ell$ is the
angular momentum quantum number, and the double braces denote of course the
integer part. An extension of this class features potentials that possess the same number of 
$\ell$-wave bound states and behave as 
$\left(a/r\right)^{2}$ both at the origin ($r\rightarrow 0^{+}$) and at
infinity ($r\rightarrow \infty$), where $a$ is an additional free parameter.
\end{abstract}

\end{frontmatter}

Recently we revisited \cite{BC1,BC2} the classical problem of establishing
upper and lower limits for the number $N_{\ell}$ of $\ell$-wave bound
states possessed by a central potential $V(r)$ (vanishing as 
$r\rightarrow \infty $) in the context of nonrelativistic quantum mechanics. As it is well
known (see, for instance, \cite{C}), this number $N_{\ell}$ coincides with
the number of zeros, in the interval $0<r<\infty$, of the zero-energy
radial wave function $u_{\ell}(r)$, namely of the solution $u_{\ell}(r)$,
characterized by the boundary condition $u_{\ell}(0)=0$, of the zero-energy
(stationary, radial) Schr\"{o}dinger equation
\begin{equation}
\label{ZeroEnSchr}
-u''_{\ell}(r)+\left[V(r)+
\frac{\ell\left(\ell+1\right)}{r^{2}}\right] \,u_{\ell}(r)=0.
\end{equation}
[Here and throughout we use of course the standard quantum-mechanical units
such that $\hbar=2m=1,$ where $m$ is the mass of the particle under
consideration, and $\ell$ (a nonnegative integer, $\ell=0,1,2,\ldots$) is the
angular momentum quantum number]. In this context we discovered a class of 
($\ell$-dependent) potentials, which (depend on two dimensionless parameters
and) have the remarkable property to possess a number $N_{\ell}$ of $\ell$
-wave bound states that (is given by a neat formula and) does not depend on
the angular momentum quantum number $\ell$. These potentials, to which we
assigned the name $V_{\text{CIC}}(r)$ because we found them while visiting
the Centro Internacional de Ciencias (CIC) in Cuernavaca (see
Acknowledgements), read as follows: 
\begin{subequations}
\label{CIC}
\begin{eqnarray} 
\label{CICa} 
V_{\text{CIC}}(g^{2},\mu^{2},\ell,R;r) &\equiv &\frac{g^{2}}{R^{2}}\,
f_{\text{CIC}}\left(\mu^{2},\ell;\frac{r}{R}\right), \\
\label{CICb}
f_{\text{CIC}}\left(\mu^{2},\ell;x\right) &=&-x^{4\ell}\,\left[ 
\left(1+\frac{x^{2\ell+1}}{2\ell +1}\right)^{2}-1+\mu^{2}\right]^{-2}.
\end{eqnarray}
\end{subequations}
Here and throughout $R,\,g^{2},\,\mu^{2}$ are three arbitrary positive
parameters, the first of which has the dimensions of a length, and
the other two are dimensionless and must satisfy the inequality 
$g^{2}+\mu^{2}>1$ (which is necessary for the existence of bound states). The
corresponding number $N_{\ell}^{\text{(CIC)}}$ of $\ell$-wave bound states
is indeed given by the following ($\ell$-independent !) neat formula
(proven below): 
\begin{equation}
N_{\ell }^{\text{(CIC)}}\left( g^{2},\mu^{2}\right) =\left\{ \left\{ 
\tilde{N}_{\ell}^{\text{(CIC)}}\left(g^{2},\mu ^{2}\right) \right\} \right\},
\label{N}
\end{equation}
where the double braces denote of course the integer part and 
\begin{subequations}
\label{NCIC}
\begin{eqnarray}
\label{NCICa}
\tilde{N}_{\ell}^{\text{(CIC)}}\left(g^{2},\mu^{2}\right) =\frac{1}
{2\pi }\sqrt{g^{2}+\mu^{2}-1}\,\left( \sqrt{1-\mu^{2}}\right)^{-1}\log 
\left[\frac{1+\sqrt{1-\mu^{2}}}{1-\sqrt{1-\mu^{2}}}\right]\\ \nonumber \quad \text{if}
\quad \mu^{2}\leq 1,
\end{eqnarray}
\begin{equation}
\label{NCICb}
\tilde{N}_{\ell}^{\text{(CIC)}}\left(g^2,\mu^2 \right) =\frac{g}{\pi}\quad 
\text{if}\quad \mu^{2}=1,
\end{equation}
\begin{eqnarray}
\label{NCICc}
\tilde{N}_{\ell}^{\text{(CIC)}}\left(g^2,\mu^2 \right) =\frac{1}{\pi}\,
\sqrt{g^{2}+\mu^{2}-1}\,\left( \sqrt{\mu ^{2}-1}\right)^{-1}\arctan 
\left(\sqrt{\mu^{2}-1}\right)\\ \nonumber \quad \text{if}\quad \mu^{2}\geq 1.
\end{eqnarray}
\end{subequations}
This remarkable property evokes a mathematical and pedagogical interest, and
it is moreover of some applicative relevance because these potentials can be
used as comparison potentials: clearly any potential $V(r)$ that is more
``attractive" than an $\ell$-wave CIC
potential, see (\ref{CIC}), $V(r)\leq V_{\text{CIC}}(g^{2},\mu ^{2},
\ell,R;r)$ for $0\leq r<\infty ,$ shall possess a number $N_{\ell}$ of $\ell$
-wave bound states at least as large as $N_{\ell }^{\text{(CIC)}}
\left(g^{2},\mu^{2}\right) $, $N_{\ell}\geq N_{\ell }^{\text{(CIC)}}
\left(g^{2},\mu^{2}\right) $, and conversely any potential $V(r)$ that is less
``attractive" than an $\ell $-wave CIC
potential, see (\ref{CIC}), $V(r)\geq V_{\text{CIC}}(g^{2},\mu^{2},\ell
,R;r)$ for $0\leq r<\infty ,$ shall possess a number $N_{\ell }$ of 
$\ell$-wave bound states not larger than $N_{\ell }^{\text{(CIC)}}\left(g^{2},
\mu^{2}\right)$, $N_{\ell }\leq N_{\ell }^{\text{(CIC)}}\left(g^{2},
\mu^{2}\right)$ \cite{BC2}. This motivated us to write the present paper, in
order to advertise this finding and to elaborate on it by providing some
information on the shape $f_{\text{CIC}}\left(\mu^{2},\ell;x\right)$ of
these potentials, see (\ref{CICb}).

First of all let us note the following qualitative features. Clearly this
function is negative, $f_{\text{CIC}}\left(\mu^{2},\ell;x\right)<0,$ for
all positive values of $x$, $0<x<\infty$, and it vanishes proportionally to
the inverse $\left[4\left(\ell+1\right)\right]$-power of $x$ at large 
$x$, 
\begin{equation}
\lim_{x\rightarrow \infty}\left[x^{4(\ell +1)}\,
f_{\text{CIC}}\left(\mu^{2},\ell;x\right)\right]=-\left(2\ell+1\right)^{4}.
\end{equation}

For $\ell=0$, this function has its minimum at $x=0$, 
\begin{equation}
\min_{0\leq x<\infty }\left[f_{\text{CIC}}\left(\mu^{2},0;x\right) 
\right]=f_{\text{CIC}}\left(\mu^{2},0;0\right) =-\mu^{-4},
\end{equation}
and it increases monotonically from this minimum value to its vanishing
value at $x=\infty $ (namely, $f'_{\text{CIC}}
\left(\mu^{2},0;x\right) >0$ for $0<x<\infty $; more specifically, if $0<\mu^{2}<6$
the second derivative of $f_{\text{CIC}}\left(\mu^{2},0;x\right)$ is
everywhere negative, $f''_{\text{CIC}}
\left(\mu^{2},0;x\right) <0$ for $0<x<\infty$, while if $\mu^{2}>6$, the second
derivative changes sign, namely $f''_{\text{CIC}}
\left(\mu^{2},0;x\right) >0$ for $0<x<\tilde{x}(\mu^{2}),
\,f''_{\text{CIC}}\left(\mu^{2},0;x\right) <0$ 
for $\tilde{x}(\mu^{2})<x<\infty$, with $\tilde{x}(\mu ^{2})=\sqrt{\frac{\mu^{2}-1}{5}}-1$).

For positive $\ell$, $\ell =1,2,\ldots$, this function vanishes at the origin
proportionally to $x^{4\ell }$ and it has a single minimum at 
$x=x_{\text{min}}\left(\mu^{2},\ell \right)$,
\begin{equation}
\label{Xmin}
x_{\text{min}}\left( \mu^{2},\ell \right) =\left[ 
\frac{\left(2\ell+1\right)}{2\left(\ell+1\right)}\,\xi(\mu^{2})
\right]^{\frac{1}{2\ell+1}},
\end{equation}
where it attains the value
\begin{eqnarray}
\min_{0\leq x<\infty }\left[ f_{\text{CIC}}\left( \mu ^{2},\ell
;x\right) \right]  &=&f_{\text{CIC}}\left[ \mu ^{2},\ell;
x_{\text{min}}\left(\mu^{2},\ell \right) \right]   \nonumber \\
\label{MinF}
&=&-\left( \ell +1\right)^{2}\left[ \frac{2(\ell +1)}{2\ell +1}
\right]^{\frac{2}{2\ell +1}}\frac{\left[\xi(\mu^{2})\right]^{\frac{4\ell }
{2\ell+1}}}{\left[ \xi (\mu ^{2})+2\mu ^{2}(\ell +1)\right] ^{2}}.
\end{eqnarray}
Here we use the short-hand notation $\xi \left( \mu ^{2}\right) =
\sqrt{1+4\ell\left(\ell+1\right)\mu^{2}}-1$. Note that $x_{\text{min}}
\left(\mu^{2},\infty \right) =1$, and $x_{\text{min}}\left(\mu^{2},\ell \right)$ 
grows as $\left[\frac{\ell \,(2\ell +1)^{2}\,\mu^{2}}
{\left(\ell+1\right) }\right]^{\frac{1}{2(2\ell +1)}}$ at large $\mu$; while the
minimum of $f_{\text{CIC}}\left(\mu ^{2},\ell ;x\right)$, see (\ref{MinF}), 
decreases as $-\ell ^{2}\,\left(1+\mu \right)^{-2}$ at large $\ell$,
and increases as $-L(\ell )\,\mu^{\frac{-4(\ell +1)}{2\ell +1}}$ at large 
$\mu$ with $L(\ell )=\left( \ell \right)^{\frac{2\ell }{2\ell +1}}
\left(\ell+1\right)^{\frac{2(\ell+1)}{2\ell+1}}$
$\left(2\ell+1\right)^{\frac{-2}{2\ell +1}}$.

The formulas (\ref{Xmin}) and (\ref{MinF}), as well of course as the
expression of the number of bound states, see (\ref{NCICb}), simplify
considerably for $\mu ^{2}=1$:
\begin{equation}
x_{\text{min}}\left(1,\ell \right) =\left[\frac{\ell 
\left(2\ell+1\right)}{\left(\ell+1\right)}\right]^{\frac{1}{2\ell +1}},
\end{equation}
\begin{equation}
\min_{0\leq x<\infty }\left[ f_{\text{CIC}}\left(1,\ell;x\right) 
\right] =f_{\text{CIC}}\left[1,\ell;x_{\text{min}}\left(1,\ell\right)\right] 
=-\ell^{\frac{4\ell}{2\ell+1}}\left(\frac{\ell+1}{2\ell+1}\right)^{\frac{4(\ell+1)}{2\ell+1}}.
\end{equation}

Another remarkable property of the function $f_{\text{CIC}}
\left(\mu^{2},\ell;x\right)$ is displayed by the formula (that can be easily
verified by direct integration) 
\begin{equation}
\label{IntF}
\int_{0}^{\infty }dx\,\sqrt{-f_{\text{CIC}}\left(\mu^{2},\ell ;x\right)}=
\frac{\pi \,g}{\sqrt{g^{2}+\mu^{2}-1}}\,
\tilde{N}_{\ell}^{\text{(CIC)}}\left(g^{2},\mu^{2}\right),
\end{equation}
see (\ref{NCIC}).

\begin{figure}
\begin{center}
\includegraphics*[width=10cm]{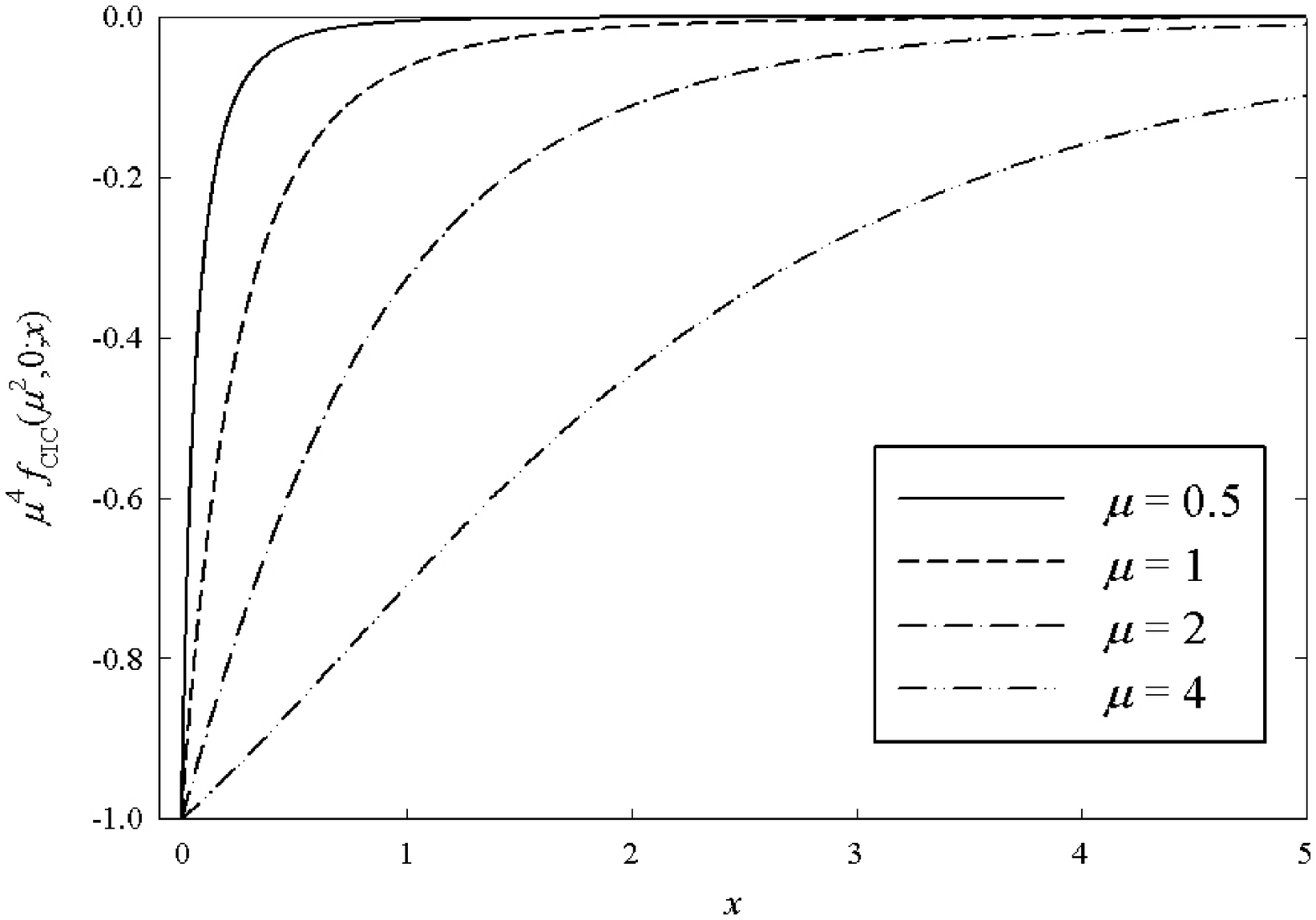}
\end{center}
\protect\caption{Graphs of the function $ \mu^4\,f_{\text{CIC}}\left(\mu^{2},0;x\right)$, see (\ref{CICb}), 
for four values of $\mu$.}
\label{fig1}
\end{figure}

\begin{figure}
\begin{center}
\includegraphics*[width=10cm]{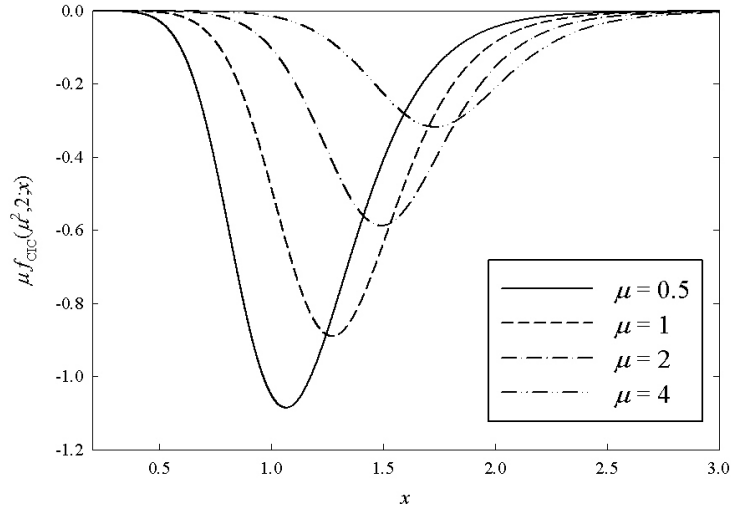}
\end{center}
\protect\caption{Graphs of the function $\mu\, f_{\text{CIC}}\left(\mu^{2},2;x\right)$, see (\ref{CICb}), 
for four values of $\mu$.}
\label{fig2}
\end{figure}

Graphs of the function $f_{\text{CIC}}\left(\mu^{2},\ell;x\right)$ (appropriately renormalized) are
presented in Figs. \ref{fig1} and \ref{fig2} for $\ell=0$ and $\ell=2$ and for various
values of $\mu$.

Let us now prove the claim made above, namely the validity of (\ref{N}) with
(\ref{NCIC}). This is done (albeit without providing here any explanation of
how this result was discovered \cite{BC2}) by relating as follows the
solution $u_{\ell }(r)$ of the zero-energy radial Schr\"{o}dinger equation 
(\ref{ZeroEnSchr}) (or rather its logarithmic derivative) to a
``phase function" $\eta_{\ell}(r)$ that
provides a convenient tool to count the zeros of $u_{\ell}(r)$:
\begin{equation}
\sqrt{-V(r)}\cot\left[\eta_{\ell}(r)\right] -\frac{V'(r)}
{4\,V(r)}=\frac{u'_{\ell}(r)}{u_{\ell}(r)}.
\end{equation}%
It is indeed plain that the boundary condition $u_{\ell}(0)=0$ entails 
$\eta_{\ell}(0)=0,$ while the definition $\eta_{\ell}(\infty)=
\tilde{N}_{\ell }\,\pi$ entails that the number $N_{\ell}$ of zeros of 
$u_{\ell}(r)$ in $0<r<\infty$ is given by the integer part of $\tilde{N}_{\ell }$. 
It is moreover easy to verify that the zero-energy radial Schr\"{o}dinger
equation (\ref{ZeroEnSchr}) yields for $\eta_{\ell}(r)$ the simple
first-order ODE
\begin{equation}
\label{ODEeta}
\eta'_{\ell}(r)\,\left\{1+\frac{\mu^{2}-1}{g^{2}}\left[ 
\sin\,\eta _{\ell}(r)\right]^{2}\right\}^{-1}=\sqrt{-V(r)},
\end{equation}
provided the potential $V(r)$ satisfies the nonlinear ODE
\begin{equation}
\label{ciceq}
-\frac{\ell\left(\ell +1\right)}{r^{2}}+\frac{5}{16}\left( 
\frac{V'(r)}{V(r)}\right)^{2}-\frac{V''(r)}{4V(r)}+
\frac{\mu^{2}-1}{g^{2}}\,V(r)=0.
\end{equation}
And it is as well easy to verify that the potential 
$V_{\text{CIC}}(g^{2},\mu^{2},\ell ,R;r)$, see (\ref{CIC}), does indeed 
satisfy this ODE, (\ref{ciceq}).

It is then plain, by taking advantage of the boundary values of 
$\eta_{\ell}(r)$ at $r=0$ and at $r=\infty$ (see above) and of the relation 
(\ref{IntF}) with (\ref{CIC}), that the integration from $r=0$ to $r=\infty$ of 
the ODE (\ref{ODEeta}) yields the expression (\ref{NCIC}) for 
$\tilde{N}_{\ell}^{\text{(CIC)}}$. Q.E.D.

\begin{figure}
\begin{center}
\includegraphics*[width=10cm]{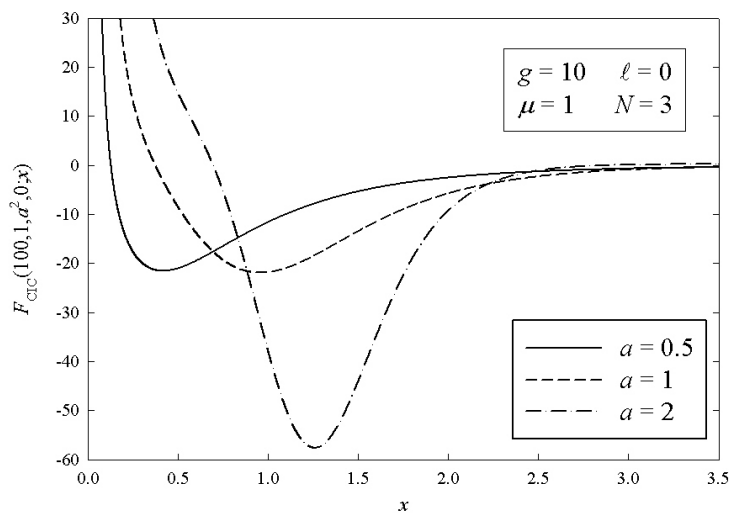}
\end{center}
\protect\caption{Graphs of the function $F_{\text{CIC}}\left(100,1,a^{2},0;x\right)$, see (\ref{Wb}), 
for three values of $a$. All these potentials, see (\ref{Wa}), possess three S-wave bound states, see (\ref{NCICb}).}
\label{fig3}
\end{figure}

\begin{figure}
\begin{center}
\includegraphics*[width=10cm]{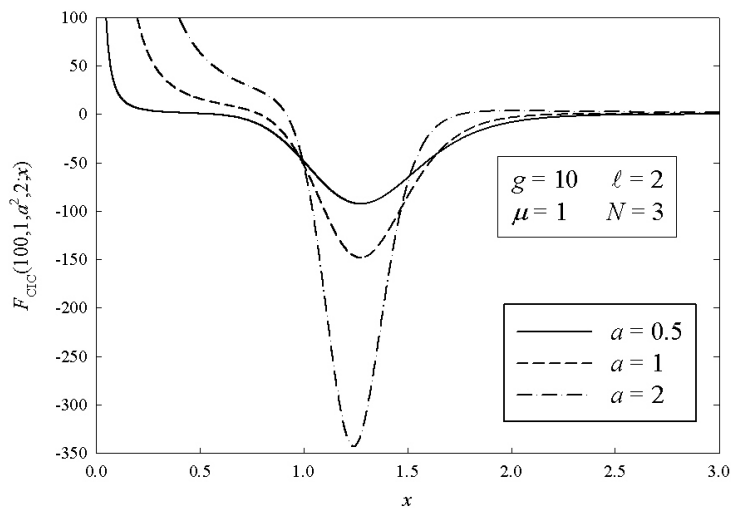}
\end{center}
\protect\caption{Graphs of the function $F_{\text{CIC}}\left(100,1,a^{2},2;x\right)$, see (\ref{Wb}), 
for three values of $a$. All these potentials, see (\ref{Wa}), possess three D-wave bound states, see (\ref{NCICb}).}
\label{fig4}
\end{figure}

\begin{figure}
\begin{center}
\includegraphics*[width=10cm]{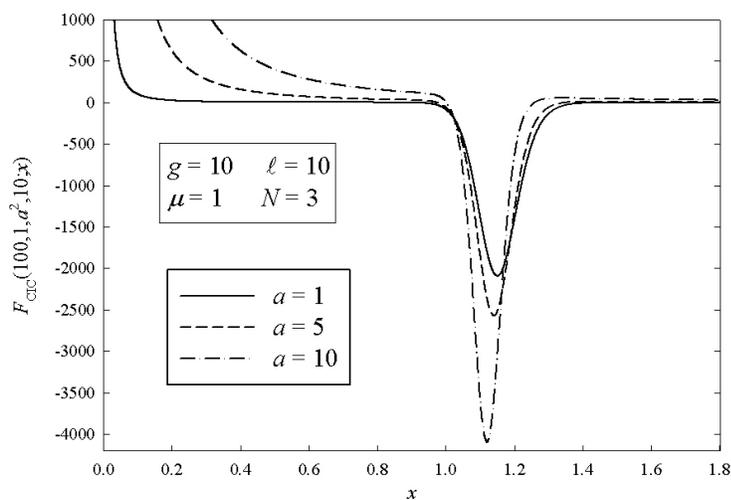}
\end{center}
\protect\caption{Graphs of the function $F_{\text{CIC}}\left(100,1,a^{2},10;x\right)$, see (\ref{Wb}), 
for three values of $a$. All these potentials, see (\ref{Wa}), possess three $\ell$-wave bound states
with $\ell=10$, see (\ref{NCICb}).}
\label{fig5}
\end{figure}

\begin{figure}
\begin{center}
\includegraphics*[width=10cm]{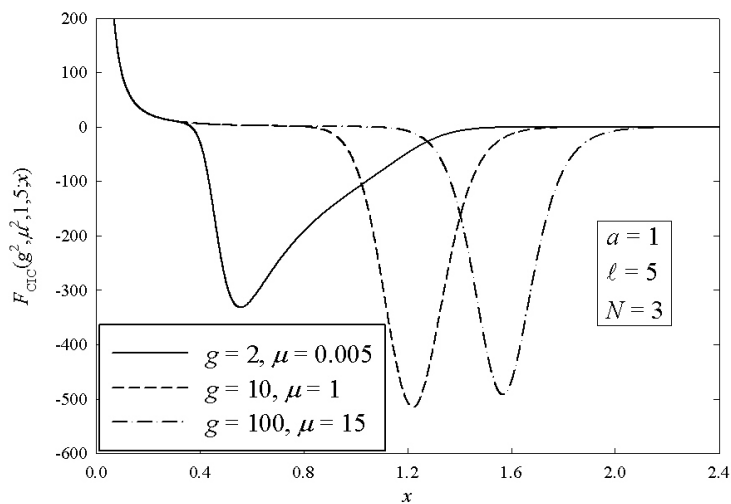}
\end{center}
\protect\caption{Graphs of the function $F_{\text{CIC}}\left(g^2,\mu^2,1,5;x\right)$, see (\ref{Wb}), 
for three values of $g$ and $\mu$ possessing three $\ell$-wave bound states with $\ell=5$, see (\ref{NCIC}).}
\label{fig6}
\end{figure}

Finally let us point out a consequence of the obvious identity of the
zero-energy Schr\"{o}dinger equation which obtains from (\ref{ZeroEnSchr})
by replacing in it $V(r)$ with $W(r),$ with the zero-energy Schr\"{o}dinger
equation which obtains from (\ref{ZeroEnSchr}) by replacing in it $\ell$
with $\lambda$, provided 
\begin{equation}
W(r)=\left(\frac{a}{r}\right)^{2}+V(r),
\end{equation}
\begin{equation}
\label{landa}
\lambda \equiv \lambda \left(a^{2},\ell \right) =-\frac{1}{2}+
\sqrt{\left(\ell+\frac{1}{2}\right)^{2}+a^{2}}. 
\end{equation}
This of course entails that the number $N_{\ell}$ of $\ell$-wave bound
states possessed by the potential 
\begin{subequations}
\label{W}
\begin{eqnarray}
\label{Wa}
W_{\text{CIC}}(g^{2},\mu^{2},a^{2},\ell,R;r) &\equiv& \frac{1}{R^{2}}\,
F_{\text{CIC}}\left(g^{2},\mu^{2},a^{2},\ell;\frac{r}{R}\right)
\\
\label{Wb}
F_{\text{CIC}}\left(g^{2},\mu^{2},a^{2},\ell;x\right) &=&
\left(\frac{a}{x}\right)^{2}+g^{2}\,f_{\text{CIC}}(\mu^{2},\lambda;x),  
\end{eqnarray}
\end{subequations}
where $a$ is an arbitrary real constant and 
$f_{\text{CIC}}(\mu^{2},\lambda ;x)$ is of course defined by 
(\ref{CICb}) with (\ref{landa}),
is still given by (\ref{N}) with (\ref{NCIC}) (this formula provides now the
number of bound-states with angular momentum quantum number $\lambda$
rather than $\ell$; but this number does not depend on this quantum number,
see (\ref{NCIC}); which of course also entails that $\lambda$ is not
required here to be an integer). This remark entails that the potential 
$W_{\text{CIC}}(g^{2},\mu^{2},a^{2},\ell,R;r)$, which depends now on the 
\textit{three }dimensionless (arbitrary, positive) constants 
$g^{2},\mu^{2},a^{2}$, provides an additional, more flexible tool to assess, by
comparison techniques, the number $N_{\ell}$ of $\ell$-wave bound states
possessed by a given central potential $V(r)$. Graphs of the function 
$F_{\text{CIC}}\left(g^{2},\mu^{2},a^{2},\ell;x\right)$ are presented in
Figs. \ref{fig3}, \ref{fig4} and \ref{fig5} for $g=10,$ $\mu=1$, $\ell=0$, $\ell=2$ 
and $\ell=10$ and for various values of $a$. In Fig. \ref{fig6}, graphs of the
function $F_{\text{CIC}}\left(g^{2},\mu^{2},a^{2},\ell;x\right)$ are presented for
$a=1$, $\ell=5$ and for various values of $g$ and $\mu$ such as the number of $\ell$-wave
bound states is always equal to three, see (\ref{NCIC}).

Let us end this paper by emphasizing that, because of the availability of
two, or even three, free parameters (for every given value of the angular
momentum quantum number $\ell$), the family of CIC potentials considered herein
is quite flexible (as also shown by the figures); these makes these
potentials -- which have a known number of bound states -- quite convenient
to be used as comparison potentials, whenever one wishes to assess quickly
and easily the number of ($\ell$-wave) bound states associated with a given
potential.

{\bf Acknowledgments}

The results reported in this paper have been obtained during the Scientific
Gathering on Integrable Systems held from November 3rd to December 13th,
2002, at the Centro Internacional de Ciencias (CIC) in Cuernavaca. It is a
pleasure to thank professor Thomas Seligman, director of CIC, and his
collaborators and colleagues for the pleasant hospitality and the fruitful
working environment provided by CIC.

\end{document}